\documentclass[aip,prd,showpacs,nofootinbib]{revtex4}
\usepackage{latexsym}
\usepackage{amsfonts}
\usepackage{amsbsy}
\usepackage{mathrsfs}

\begin{document}

\title{Minimally modified self-dual 2-forms gravity}

\date{\today}

\author{Riccardo Capovilla}\email{capo@fis.cinvestav.mx}

\author{Merced Montesinos}\email{merced@fis.cinvestav.mx}

\author{Mercedes Vel\'azquez}\email{mquesada@fis.cinvestav.mx}

\affiliation{ Departamento de F\'{\i}sica, Cinvestav, Instituto Polit\'ecnico
Nacional 2508, San Pedro Zacatenco, 07360, Gustavo A. Madero, Ciudad de
M\'exico, M\'exico.}

\begin{abstract}
The first order Pleba\'{n}ski formulation of (complex) general relativity (GR)
in terms of self-dual 2-forms admits a generalization, proposed by Krasnov,
that is qualitatively different from other possible generalizations of GR in
terms of metric variables. In this paper, we investigate, within a minimal
modification, and in a perturbative approach, the geometrical meaning of the
field variables used in the Krasnov generalization, and compare them to the
field variables used in the Pleba\'{n}ski formulation.
\end{abstract}


\maketitle

About a century ago, Einstein discovered that the basic field variables for
gravity were given by the spacetime metric, in his theory of GR. After that,
various formulations of GR have been proposed. Cartan proposed an alternative
formulation in  terms of a tetrad $\theta$, the `square root' of the metric,
$g = \theta \otimes \theta$. Another formulation, of interest for the purposes
of this paper, was introduced in the mid seventies by Pleba\'{n}ski
\cite{Pleb}, and further clarified and expanded to include couplings to matter
in \cite{CDJM}, where it is shown that it is the natural covariant formulation
of the Ashtekar Hamiltonian formulation of (complex) GR \cite{1Ashtekar}. The
basic field variables are  a triplet of self-dual 2-forms $\Sigma^{AB}$, that
is surfaces
\begin{equation}
\Sigma^{AB} = \frac12 \theta^{AA'} \wedge \theta^B{}_{A'} \,,
\label{eq:sigma}
\end{equation}
where $\theta^{AA'}$ is a tetrad in the Cartan formalism, and we follow the
notation of \cite{gerardo}. The first-order Pleba\'{n}ski action for (complex)
vacuum GR is of the $BF$ form with an additional constraint:
\begin{equation}
S [ {\cal A}, \Sigma, \Psi ] =
\int  {\cal F}_{AB} \wedge \Sigma^{AB} - {1\over 2} \Psi_{ABCD} \; \Sigma^{AB} \wedge
\Sigma^{CD}\,,
\label{eq:pl}
\end{equation}
where  ${\cal F}_{AB} = d {\cal A}_{AB} - {\cal A}_{AC}  \wedge {\cal
A}_B{}^{C}$ is the curvature of the $SL(2,\mathbb{C})$ connection ${\cal
A}^{AB}$. The Lagrangian multiplier $\Psi_{ABCD} = \Psi_{(ABCD)}$ is totally
symmetric. This elegant first-order formulation of (complex) GR implies the
following equations of motion:
\begin{eqnarray}
&& \delta \Psi: \Sigma^{(AB} \wedge \Sigma^{CD)} = 0\,, \label{eq:ss} \\
&& \delta {\cal A}: {\cal D} \Sigma ^{AB} = d \Sigma^{AB} - 2
{\cal A}^{(A}{}_C \wedge  \Sigma^{B)C} = 0\,, \label{eq:3}\\
&& \delta \Sigma: {\cal F}_{AB} =   \Psi_{ABCD} \; \Sigma^{CD} \,.
\label{eq:e}
\end{eqnarray}
The first ensures that $\Sigma^{AB}$ has the form (\ref{eq:sigma}). The second
identifies ${\cal A}^{AB}$ with the self-dual spin connection, i.e. the
self-dual part of the torsion-less spin connection $\omega^{AA'}{}_{BB'}$
compatible with $\theta^{AA'}$, given by  $d\theta^{AA'} -
\omega^{AA'}{}_{BB'} \wedge \theta^{BB'} = 0$, with $\omega_{AA'BB'} =
\epsilon_{A'B'} {\cal A}_{AB} + \epsilon_{AB} {\cal A}_{A'B'}$, where ${\cal
A}_{A'B'}$ is the anti-self-dual part. Consequently, ${\cal F}_{AB}$ is the
self-dual part of the spacetime curvature. The third equation expresses the
vacuum Einstein equations in a somewhat unusual form: the self-dual curvature
is given  purely in terms of the Lagrange multiplier $\Psi_{ABCD}$. This
identifies $\Psi_{ABCD}$ with the Weyl spinor. Since, in four dimensions, the
curvature is the sum of the Weyl part plus the Ricci part, this is equivalent
to the statement that the Ricci part of the curvature vanishes, that is
Einstein's equations in vacuum for complex GR. For a spacetime of Lorentzian
signature, the field variables are complex. In order to obtain a real solution
of Einstein's equations one needs to impose reality conditions on the field
variables.

The introduction of a cosmological constant $\Lambda$ does not change
the basic structure:
\begin{equation}
S [ {\cal A}, \Sigma, \Psi ] =
\int  {\cal F}_{AB} \wedge \Sigma^{AB} - {1\over 2} \Psi_{ABCD} \; \Sigma^{AB} \wedge
\Sigma^{CD} - \frac{\Lambda}{6} \; \Sigma^{AB} \wedge \Sigma_{AB} \,.
\label{eq:pll}
\end{equation}
The extra term involving $\Lambda$ is proportional to the spacetime
volume element. For the field equations,
the only modification is in (\ref{eq:e}), that takes the form
\begin{equation}
 {\cal F}_{AB} = \Psi_{ABCD} \Sigma^{CD}  + \frac{\Lambda}{3} \Sigma_{AB}\,.
 \label{eq:LL}
\end{equation}


Recently, Krasnov has proposed a modification of the Pleba\'{n}ski action that
is the subject of this paper \cite{kr1,kr2,kr3,kr4} (see also
\cite{lee,ben,lf}). The basic idea is to turn the cosmological constant into a
function by modifying the constraint that appears in the action (\ref{eq:pl}).
Krasnov considers the modified action
\begin{equation}
S [ A, B , \phi] =
\int  F_{AB} \wedge B^{AB} - {1\over 2} \phi_{ABCD} \; B^{AB} \wedge
B^{CD} - \frac12 \Phi (\phi_2, \phi_3 ) B^{AB} \wedge B_{AB}\,,
\label{eq:kr}
\end{equation}
where $B^{AB}$ is a triplet of 2-forms, the totally symmetric spinor
$\phi_{ABCD} = \phi_{(ABCD)}$ would be the analog of $\Psi_{ABCD}$ in
(\ref{eq:pl}), and has been called inappropriately the Weyl spinor. It has in
common with $\Psi_{ABCD}$ that they both are Lagrange multipliers and totally
symmetric. $\Phi (\phi_2, \phi_3)$ is an arbitrary function of the only two
independent algebraic invariants for the totally symmetric $\phi_{ABCD}$:
$\phi_2 = \phi^{ABCD} \phi_{ABCD}$, and $\phi_3 = \phi^{ABCD} \phi_{CDEF}
\phi^{EF}{}_{AB}$. $F_{AB} = d A_{AB} - A_{AC}
 \wedge A_B{}^{C}$  is the curvature of the  $SL(2,\mathbb{C})$ connection
 $A_{AB}$. We avoid including a cosmological constant  $\Lambda$ term, since
 it can be included as a constant term in $\Phi (\phi_2, \phi_3 )$. We emphasize that
 all the field variables are valued in  $SL(2,\mathbb{C})$.

This modified action defines a class of generally covariant theories that
reduces to vacuum (complex) GR in the formulation (\ref{eq:pl}) when $\Phi \to
0$. There are additional fields that enter in an `economical' way: as shown by
Krasnov \cite{kr3}, a Hamiltonian analysis of (\ref{eq:kr}) shows that in this
class there are two propagating degrees of freedom, just like GR. The
formulation of GR in terms of 2-forms therefore admits a generalization that
is qualitatively different from the generalizations of GR proposed and widely
explored in terms of metric variables that involve higher powers of the
curvature and imply in general the addition of extra degrees of freedom.
However, the consequences of the Krasnov modification to the Pleba\'{n}ski
action imply a radical change in the basic geometric structure that underlies
the Pleba\'nski formulation. In particular, we are interested in understanding
the geometric meaning of the field variables $\{A, B , \phi \}$. This paper
tries to contribute an epsilon, as we illustrate below.

 We focus on a minimal version of the class of theories considered
 by Krasnov, specializing from the outset to  the choice
 \begin{equation}
  \Phi (\phi_2, \phi_3 ) = -\frac{\varepsilon}{2}\;
  \phi^{ABCD} \phi_{ABCD}\,,
  \label{eq:var2}
\end{equation}
where we have introduced a numerical parameter $\varepsilon$ in order to
quantify the modification from (\ref{eq:pl}). Most of our considerations do
extend to the general case (\ref{eq:kr}), but we think it is useful to look
first at the simplest possible example.

The equations of motion that follow from this minimally modified action are as
follows
\begin{eqnarray}
&&\delta \phi: \quad B^{(AB} \wedge B^{CD)} =  \varepsilon  \phi^{ABCD} B^{EF} \wedge B_{EF}\,, \label{eq:beta}\\
&& \delta A: \quad D B^{AB} = d B^{AB} - 2  A^{(A}{}_C \wedge  B^{B)C} = 0\,, \label{eq:A}\\
&& \delta B: \quad F_{AB} = \phi_{ABCD} B^{CD}  -  \frac12 \varepsilon\, \phi_2  B_{AB}\,. \label{eq:f}
\end{eqnarray}
The introduction of a non-trivial non-vanishing rhs in the constraint
(\ref{eq:beta}) ruins the beautiful orthogonality condition  (\ref{eq:ss}).
Moreover, it introduces priviledged directions, along the eigenspinors of
$\phi^{ABCD}$, that can be classified according to their algebraic type, just
like the Petrov classification of the Weyl spinor. This is an important point,
as the internal $SL(2,\mathbb{C})$ symmetry is inevitably broken,  that we
plan to explore in future work \cite{CMM1}. Now we do not know what is the
geometrical meaning  of the triplet of 2-forms $B^{AB}$. In turn, this
modification of the Pleba\'{n}ski action obscures the geometrical meaning of
the connection $A_{AB}$, and consequently of its curvature $F_{AB}$. Since, at
this stage, we do not know the geometrical meaning of the curvature $F_{AB}$,
to identify $\phi_{ABCD}$ with the Weyl spinor $\Psi_{ABCD}$ is clearly
premature.

In order to understand the geometric meaning of the field variables $\{A, B ,
\phi \}$, we expand them as follows:
\begin{eqnarray}
B^{AB} &=& \Sigma^{AB}  + \varepsilon \sigma^{AB}\,, \label{eq:b}\\
A^A{}_B &=& {\cal A}^A{}_B + \varepsilon \alpha^A{}_B\,, \label{eq:a} \\
\phi_{ABCD} &=& \Psi_{ABCD} + \varepsilon \rho_{ABCD}\,, \label{eq:r}
\end{eqnarray}
where the set of field variables  $\{ \Sigma, {\cal A}, \Psi \}$ satisfies the
Einstein vacuum equations (\ref{eq:ss})-(\ref{eq:e}), and the set of field
variables $\{ \sigma, \alpha, \rho\}$ are `corrections'. Note that we are
using the same parameter $\varepsilon$ that appears in the term
(\ref{eq:var2}). {\it A priori}, there is no reason why they should be the
same. This is an arbitrary yet natural assumption, that perhaps ought to be
relaxed. We emphasize that this $\varepsilon$-expansion is to be understood in
the space of the class of generally covariant theories defined by
(\ref{eq:kr}), with the Pleba\'nski formulation (\ref{eq:pl}) as the `origin'.
At this point, no physical meaning can be attached to the parameter
$\varepsilon$. It is just a working hypothesis.

Inserting  the expansions (\ref{eq:b})-(\ref{eq:r}) into the field equations
(\ref{eq:beta})-(\ref{eq:f}), and keeping only terms up to order
$\varepsilon$, assuming of course that $\varepsilon$ is small, we obtain
\begin{eqnarray}
2 \Sigma^{(AB} \wedge \sigma^{CD)} &=& \Psi^{ABCD} \Sigma^{EF} \wedge \Sigma_{EF}\,, \label{eq:ss1} \\
2 \alpha^{(A}{}_C \wedge \Sigma^{B) C} &=& {\cal D}  \sigma^{AB}\,, \label{eq:A1} \\
{\cal D} \alpha_{AB} &=& \rho_{ABCD} \Sigma^{CD} + \Psi_{ABCD}\, \sigma^{CD} -
\frac12 \Psi_2 \Sigma_{AB}\,, \label{eq:da}
\end{eqnarray}
where $\Psi_2$ is the invariant constructed out of the Weyl spinor, $\Psi_2=
\Psi_{ABCD}\Psi^{ABCD}$.

Our task is now to understand the geometrical meaning of the  corrections
 $\{ \sigma , \alpha , \rho \}$ as follows from these equations.
 We begin with $\sigma$. Since it is a 2-form, we can expand it with respect to the
basis $\{ \Sigma^{AB} , \Sigma^{A'B'} \} $  that span the
 space of 2-forms, where $\Sigma^{A'B'} = (1/2) \theta^{CA'}
 \wedge \theta_C\,^{B'}$ are a triplet of anti-self-dual 2-forms.
 We recall that they are orthogonal to $\Sigma^{AB}$:
  \begin{equation}
 \Sigma^{AB} \wedge \Sigma^{A'B'} = 0\,.
 \end{equation}
 We will also use the identity
 \begin{equation}
 \Sigma^{AB} \wedge \Sigma^{CD} = - {1 \over 3} \epsilon^{A(C} \epsilon^{D)B}
 \; \Sigma^{EF} \wedge \Sigma_{EF}\,.
 \end{equation}
 We have then that, expanding in components,
 \begin{equation}\label{esigma}
 \sigma^{AB} = \sigma^{AB}{}_{CD} \Sigma^{CD} + \sigma^{AB}{}_{C'D'} \Sigma^{C'D'}\,.
 \end{equation}
Now, the constraints (\ref{eq:ss1}) are 5 equations for the 18 components of
$\sigma^{AB}$; therefore, the solution must involve 13 free parameters. By
plugging (\ref{esigma}) into
 (\ref{eq:ss1}), we see that it implies
\begin{eqnarray}
\sigma^{AB} &=& {3 \over 2} \Psi^{ABCD} \Sigma_{CD}   +
\kappa^{(A}{}_C \Sigma^{B)C} + \sigma^{AB}{}_{C'D'} \Sigma^{C'D'}\,, \label{s1}\\
& = & {3 \over 2} {\cal F}^{AB}   + \chi^{(A}{}_C \Sigma^{B)C} + \kappa \Sigma^{AB}+
\sigma^{AB}{}_{C'D'} \Sigma^{C'D'}\label{s2}\,.
\end{eqnarray}
The first term is a shift of $\Sigma^{AB}$ along the self-dual curvature. The
specific form of this particular term depends on our special choice
(\ref{eq:var2}) for the function $\Phi (\phi_2, \phi_3) $. The four components
of $\kappa^{AB}$ are the novel fields in the self-dual part of $\sigma$,
completely undetermined. It turns out to be convenient to split $\kappa^{AB}$
in its symmetric and anti-symmetric parts:
\begin{equation}
\kappa^{AB} = \chi^{AB} + \epsilon^{AB} \kappa\,,
\end{equation}
with $\chi^{AB} = \chi^{(AB)}$ and $\kappa = (1/2) \epsilon^{AB} \kappa_{AB}$.
The last term in (\ref{s1}) implies that in general $\sigma^{AB}$   possesses
a non-vanishing anti-self-dual part, $\sigma^{AB}\,_{A'B'}$, again
undetermined. To use Pleba\'{n}ski's cherished terminology \cite{Pleb}
(apparently first introduced by Newman \cite{newman}), $B^{AB}$ is an {\it
earthly} object (self-dual: {\it heavenly}, anti-self-dual: {\it hellish},
both: {\it earthly}). Although our analysis is limited by our approximation to
first order in $\varepsilon$, this feature does not depend on our special
choice (\ref{eq:var2}) for the function $\Phi$. In general therefore Krasnov's
modification implies a mix of the self-dual and anti-self-dual parts of the
space of 2-forms where self-duality is of course understood with respect to
the $\varepsilon=0$ point.

Now, let us consider (\ref{eq:A1}) and (\ref{eq:da}). First, we note that by
taking the (self-dual) covariant derivative of  (\ref{eq:A1}) we get
\begin{eqnarray}\label{consistencia}
-{\cal D} \alpha^{\left ( A \right.}\,_C \wedge \Sigma^{B \left. \right )C} =
{\cal F}^{\left ( A \right.}\,_C \wedge \sigma^{B \left. \right )C}\,.
\end{eqnarray}
By inserting into this equation the expressions for ${\cal D} \alpha^{AB}$
given in  (\ref{eq:da}), $\sigma^{AB}$ given in (\ref{s1}), and ${\cal
F}^{AB}$ given in (\ref{eq:e}) this equation is satisfied  identically without requiring
the knowledge of the explicit form of the correction $\rho_{ABCD}$ to the Weyl
spinor. 
In this sense, (\ref{eq:A1}) and (\ref{eq:da}) are consistent with each other.
It is worth noting that this compatibility holds also when the anti-self-dual
part of $\sigma^{AB}$ is non vanishing.

The next logical step is to consider  (\ref{eq:A1}) in order to determine
$\alpha^{AB}$; these are  12 equations for the 12 components of $\alpha^{AB}$;
the solution is unique. Since $\alpha^{AB}$ is a trio of 1-forms, we expand it
with respect to the tetrad $\theta^{AA'}$ as $\alpha^{AB} =
-\alpha^{AB}{}_{DD'} \theta^{DD'}$. Then  (\ref{eq:A1}) becomes
\begin{equation}\label{eq:A1:1}
-2 \alpha^{(A}{}_{CDD'}  \Sigma^{B) C} \wedge \theta^{DD'} = {\cal D}  \sigma^{AB}\,.
\end{equation}
It is useful to introduce a set $\{ {\check{\theta}}^{AA'} \}$  of four
linearly independent 3-forms that satisfy the relations
\begin{equation}\label{cuernos}
\Sigma^{AB} \wedge \theta^{CC'}=\epsilon^{C(A}
{\check{\theta}}^{B)C'}\,, \quad  \Sigma^{A'B'} \wedge \theta^{CC'}=-
\epsilon^{C'(A'} {\check{\theta}}^{ C  B')}\,.
\end{equation}
Using (\ref{cuernos}), the lhs of (\ref{eq:A1:1}) acquires the form
\begin{eqnarray}
2 \alpha^{(A}{}_{CDD'}  \Sigma^{B) C} \wedge \theta^{DD'}  =
\left( \alpha^{(A}\,_E\,^{B)}\,_{E'} + \alpha^{(A}\,_C\,^{|C|}\,_{E'}
\epsilon^{B)}\,_E \right ) {\check{\theta}}^{EE'}\,.
\end{eqnarray}
The rhs of (\ref{eq:A1:1}) can be  rewritten using $d f = - (\partial_{EE'} f
) \theta^{EE'}$, where $f$ is a function,  and the fact that $d \Sigma^{AB}= -
{\cal A}^{(A}\,_C\,^{B)}\,_{D'} {\check{\theta}}^{CD'}- {\cal
A}^{(A}\,_C\,^{\mid C \mid}\,_{D'} {\check{\theta}}^{B)D'}$, $d \Sigma^{A'B'}=
{\cal A}^{(A'}\,_{C'}\,^{B')}\,_D {\check{\theta}^{DC'}} + {\cal
A}^{(A'}\,_{C'}\,^{\mid C'}\,_D {\check{\theta}}^{D \mid B')}$ where ${\cal
A}_{AB}=- {\cal A}_{ABCC'} \theta^{CC'}$ and ${\cal A}_{A'B'}=- {\cal
A}_{A'B'C'C} \theta^{CC'}$, so that
\begin{eqnarray}
 {\cal D}  \sigma^{AB} = \left ( - \nabla^C\,_{E'} \sigma^{AB}\,_{CE} +
 \nabla_E\,^{C'} \sigma^{AB}\,_{C'E'}
\right ) {\check{\theta}}^{EE'},
\end{eqnarray}
with $\nabla_{DD'} \mu_{ABCC'} := \partial_{DD'} \mu_{ABCC'} + {\cal
A}_A\,^E\,_{DD'} \mu_{EBCC'} + {\cal A}_B\,^E\,_{DD'} \mu_{AECC'} + {\cal
A}_C\,^E\,_{DD'} \mu_{ABEC'} + {\cal A}_{C'}\,^{E'}\,_{D'D} \mu_{ABCE'}$.
Therefore, using the fact that ${\check{\theta}}^{AA'}$ are linearly
independent, we have
\begin{eqnarray}
\alpha_{ABCC'} &=& -\nabla^E\,_{C'} \sigma_{ABEC} + 2 \nabla^E\,_{C'}  \sigma_{C(AB)E} -
\epsilon_{C(A} \nabla^E\,_{C'} \sigma^D\,_{B)ED}\nonumber\\
&& + \nabla_C\,^{E'} \sigma_{ABE'C'} - 2 \nabla_{(A}\,^{E'} \sigma_{B)CE'C'} +
\epsilon_{C(A} \nabla_{\mid D \mid}\,^{E'} \sigma^D\,_{B)E'C'}.
\end{eqnarray}

By inserting the explicit expression for $\sigma_{ABCD}$ coming from
(\ref{s1}), i.e.  $\sigma_{ABCD}= \frac32 \Psi_{ABCD} + \frac12 \left (
\chi_{A(C} \epsilon_{\mid B \mid D)} + \chi_{B(C} \epsilon_{\mid A \mid D)}
\right ) +  \kappa \epsilon_{A(C} \epsilon_{\mid B \mid D)}$, we get
\begin{eqnarray}\label{solalpha}
\alpha_{ABCC'} &=& \frac32 \nabla^E\,_{C'} \Psi_{ABEC} + \frac12 \nabla_{CC'} \chi_{AB} +
\frac12 \epsilon_{C(A}\nabla_{B)C'} \, \kappa \nonumber\\
&& + \nabla_C\,^{E'} \sigma_{ABE'C'} - 2 \nabla_{(A}\,^{E'} \sigma_{B)CE'C'} +
\epsilon_{C(A} \nabla_{\mid D \mid}\,^{E'} \sigma^D\,_{B)E'C'}\,,
\end{eqnarray}
with $\nabla_{AA'} \kappa = \partial_{AA'} \kappa$.
This complicated expression can be simplified by noting that the first term vanishes
on account of the Bianchi identities so we have
\begin{eqnarray}\label{solalpha1}
\alpha_{ABCC'} &=&  \frac12 \nabla_{CC'} \chi_{AB} +
\frac12 \epsilon_{C(A}\nabla_{B)C'} \, \kappa \nonumber\\
&&+ \nabla_C\,^{E'} \sigma_{ABE'C'} - 2 \nabla_{(A}\,^{E'} \sigma_{B)CE'C'} +
\epsilon_{C(A} \nabla_{\mid D\mid}\,^{E'} \sigma^D\,_{B)E'C'}\,.
\end{eqnarray}

If we restrict our attention to field configurations with vanishing
anti-self-dual part, $\sigma^{AB}\,_{A'B'}=0$, we obtain the remarkably simple
expression
\begin{eqnarray}
\alpha_{ABCC'} &=&  \frac12 \nabla_{CC'} \chi_{AB} +
\frac12 \epsilon_{C(A}\nabla_{B)C'} \, \kappa \,.
\end{eqnarray}
or
\begin{eqnarray}
\alpha^{AB} =  \frac12 {\cal D} \chi^{AB} + \beta^{AB},
\end{eqnarray}
with $\beta^{AB}= \frac12 \left ( \nabla^{(A}\,_{C'} \, \kappa \right )
\theta^{B) C'} $. Notice the identity $2\beta^{(A}\,_C \wedge \Sigma^{B)C} = d \kappa \wedge \Sigma^{AB}$.

Coming back to the generic case, the next step would be to insert the
expression for (\ref{solalpha1}) into  the lhs of (\ref{eq:da}) to derive what
is $\rho_{ABCD}$. However, it turns out to be a more convenient strategy to
arrive to $\rho_{ABCD}$ directly from the field equation (\ref{eq:beta}) by
solving  for
\begin{eqnarray}
\phi^{ABCD}= \frac{1}{\varepsilon} \frac{B^{ \left( AB \right.}\,_{\alpha\beta}
B^{\left. CD \right )}\,_{\gamma\delta} \, \varepsilon^{\alpha\beta\gamma\delta}}{B^{EF}\,_{\alpha\beta}
B_{EF \gamma\delta} \, \varepsilon^{\alpha\beta\gamma\delta}}\,,
\end{eqnarray}
and  with (\ref{eq:b}) we have
\begin{eqnarray}
\phi^{ABCD}= \frac{ \left ( 2 \Sigma^{(AB}\,_{\alpha\beta} \sigma^{CD)}\,_{\gamma\delta}
+ \varepsilon  \sigma^{(AB}\,_{\alpha\beta} \sigma^{CD)}\,_{\gamma\delta} \right )
\varepsilon^{\alpha\beta\gamma\delta}}{\left ( \Sigma^{AB}\,_{\alpha\beta} \Sigma_{AB\gamma\delta}
+ 2 \varepsilon \Sigma^{AB}\,_{\alpha\beta} \sigma_{AB \gamma\delta} +
\varepsilon^2 \sigma^{AB}\,_{\alpha\beta} \sigma_{AB \gamma\delta}
\right )
\varepsilon^{\alpha\beta\gamma\delta}}.
\end{eqnarray}
Inserting (\ref{s1}), expanding the denominator around $\varepsilon=0$ and
keeping terms up to order $\varepsilon$, we derive (\ref{eq:r}) with
\begin{eqnarray}\label{finalr}
\rho^{ABCD} &=& \frac34 \Psi^{\left (AB\right.}\,_{EF} \Psi^{\left. CD \right)EF} + \chi^{\left(A\right.}\,_E
\Psi^{\left. BCD \right )E} - \frac16 \chi^{\left ( AB \right.} \chi^{\left. CD \right )} \nonumber\\
&& - \kappa \Psi^{ABCD} - \frac13 \sigma^{\left(AB \right. \mid C'D' \mid} \sigma^{\left. CD \right )}\,_{C'D'}\,.
\end{eqnarray}
A first consideration is that $\phi_{ABCD}$ is not the Weyl spinor
$\Psi_{ABCD}$, because $\rho_{ABCD}$ is in general non vanishing. In fact,
$\rho_{ABCD}$ presents an interesting structure in terms of higher order
curvature terms. We also find it interesting that the undetermined self-dual
fields $\kappa_{AB}$ appear together with the Weyl spinor. The terms quadratic
in the fields are probably an effect of our approximations and
simplifications.

In conclusion, our work has obvious limitations. Perhaps, the most important
is that we are considering only  a modification of the self-dual sector of GR.
The field variables are valued in $SL(2,\mathbb{C})$. Although this sector has
been very useful elucidating alternative structures for GR, it does imply the
complication of the need to impose reality conditions {\it a posteriori}. The
fact that the field $B^{AB}$ is {\it earthly} strongly suggests that one
should consider the real formulation of GR in terms of 2-forms where the
action is formulated in terms of field variables valued in the full Lorentz
group $SO(3,1)$ (see e.g. \cite{cqg1999,cqg1999b,real} and also
\cite{smolin}). A second point where the work presented here should be of some
help is in addressing the problem of the coupling to matter of spin 1/2 and
spin 3/2, along the lines of \cite{CDJM}. For this, a better understanding of
the geometrical content of the connection and its curvature that goes beyond
our limited approach is essential. Another subject for future work is to
unravel how the extra fields that appear in the modified theory somehow do not
contribute additional degrees of freedom. Hopefully, the exercise presented in
this paper will be useful towards clarifying some aspects of the geometrical
content of the class of generally covariant theories proposed by Krasnov.

We thank Kirill Krasnov for useful and prompt comments. This work was
supported in part by CONACYT, Mexico, grant numbers 56159-F and 128243-F.

\end{document}